\documentclass[aps,prc,twocolumn,superscriptaddress,showpacs]{revtex4}

\usepackage{graphicx}
\usepackage{dcolumn}
\usepackage{bm}
\usepackage{ulem}
\usepackage{color}

\newcommand{\al}{$\alpha$}

\newcommand{\raa}{($\alpha$,$\alpha$)}

\newcommand{\rag}{($\alpha$,$\gamma$)}
\newcommand{\ran}{($\alpha$,n)}
\newcommand{\rap}{($\alpha$,p)}

\newcommand{\stot}{$\sigma_{\rm{reac}}$}
\newcommand{\sred}{$\sigma_{\rm{red}}$}

\begin{document}

\title{Total reaction cross sections from elastic $\alpha$-nucleus scattering
  angular distributions around the Coulomb barrier
}

\author{P.\,Mohr}
\email{WidmaierMohr@t-online.de}
\affiliation{
Diakonie-Klinikum, D-74523 Schw\"abisch Hall, Germany}
\affiliation{
Institute of Nuclear Research (ATOMKI), H-4001 Debrecen, Hungary}
\author{D.\ Galaviz}
\affiliation{Centro de Fisica Nuclear, University of Lisbon, 1649-003 Lisbon,
  Portugal} 
\author{Zs.\,F\"ul\"op}
\author{Gy.\,Gy\"urky}
\author{G. G.\,Kiss}
\altaffiliation [present address:]{ Laboratori Nazionali del Sud, INFN,
  Catania, Italia} 
\author{E.\,Somorjai}
\affiliation{
Institute of Nuclear Research (ATOMKI), H-4001 Debrecen, Hungary}

\date{\today}

\begin{abstract}
The total reaction cross section $\sigma_{\rm{reac}}$ is a valuable measure
for the prediction of $\alpha$-induced reaction cross sections within the
statistical model and for the comparison of scattering of tightly bound 
projectiles to weakly bound and exotic projectiles. Here we provide the
total reaction cross sections $\sigma_{\rm{reac}}$ derived from our previously
published angular distributions of $\alpha$-elastic scattering on $^{89}$Y,
$^{92}$Mo, $^{112,124}$Sn, and $^{144}$Sm at energies around the Coulomb
barrier.
\end{abstract}


\maketitle

The total reaction cross section \stot\ is related to the complex scattering
matrix $S_L = \eta_L \, \exp{(2i\delta_L)}$ by the well-known relation
\begin{equation}
\sigma_{\rm{reac}} = \frac{\pi}{k^2} \sum_L (2L+1) \, (1 - \eta_L^2)
\label{eq:stot}
\end{equation}
where $k = \sqrt{2 \mu E}/\hbar$ is the wave number, and $\eta_L$ and
$\delta_L$ are the real reflexion coefficients and scattering phase
shifts. Usually, experimental elastic scattering angular distributions are
analyzed using a complex optical potential. Although there are considerable
uncertainties in the extraction of the optical potential from elastic
scattering at low energies around the Coulomb barrier, 
the underlying reflexion coefficients and phase shifts are well
defined, as long as the measured angular distribution covers the full angular
range from $0^\circ$ to $180^\circ$ and it is well described by the chosen
optical potential. Thus, the total reaction 
cross section \stot\ can be extracted from experimental elastic scattering
angular distributions with small uncertainties. Although the analysis uses
optical potentials, i.e.\ input from theory, in practice the obtained total
reaction cross section \stot\ can be considered an experimental quantity
because of its insensitivity to the potential as long as the potential
accurately reproduces the elastic scattering angular distribution.  

A completely model-independent determination of \stot\ is also possible by a
direct fit of the reflexion coefficients $\eta_L$ and phase shifts $\delta_L$
to the angular distribution. However, in practice such fits are not widely
used because $\eta_L$ and $\delta_L$ have to be determined for angular momenta
up to about $L = 30$, i.e.\ around 60 parameters have to be fitted
simultaneously. This may lead to an oscillatory behavior of the calculated
cross sections in angular regions where no experimental data are available to
restrict the multi-parameter model-independent fit \cite{Roos87}. Therefore,
model-independent analyses are not the ideal tool for the determination of
total reaction cross sections \stot .

It has unfortunately become a common practice in optical model analyses of \al-elastic scattering data not to give \stot\ explicitly although this number
is implicitly determined in any optical model calculation, as
Eq.~(\ref{eq:stot}) shows. Nevertheless it has been noticed recently that
\stot\ is a very valuable quantity in several respects. First, a significantly
different energy dependence of \stot\ has been observed in the comparison of
scattering of tightly bound projectiles like \al\ or $^{16}$O, weakly
bound projectiles like $^{6,7,8}$Li, and exotic projectiles with halo
properties like $^{6}$He and $^{11}$Be
(e.g.\ \cite{Pie10,Far10,Kee09,Sat91}). As a consequence, \stot\ is
practically always provided in analyses of $^6$He scattering. Second,
the cross section of \al -induced reactions like e.g.\ \rag , \ran , or
\rap\ sensitively depends on the underlying \al -nucleus potential
that is applied in statistical model calculations, e.g. for the prediction of
reaction cross sections and stellar reaction rates for nuclear astrophysics 
\cite{Avr09,Rau03b,Dem02}. In these
calculations the total reaction cross section is taken as the formation cross
section of the compound nucleus, and it is thus the basic building block for
any prediction of reaction cross sections within the statistical model. 
(Note, however, that this basic
prerequisite of the statistical model may be not fulfilled at very low
energies where direct inelastic scattering may at least contribute to \stot .)

The present Brief Report provides the total reaction cross sections
\stot\ from high precision \al\ scattering experiments performed at
ATOMKI, Debrecen, Hungary \cite{Mohr97,Ful01,Gal05,Kiss09}. These scattering
data cover an angular range between $20^\circ$ and $170^\circ$ in steps
of about $1^\circ - 2^\circ$ and show small error bars below $\approx 4$\,\%
over the whole angular range which is by far sufficient to derive the total
reaction cross section with small uncertainties. These uncertainties in the
determination of \stot\ from experimental elastic scattering
angular distributions are briefly discussed.  

\begin{figure}[htb]
\includegraphics[bbllx=15,bblly=30,bburx=440,bbury=375,width=\columnwidth,clip=]{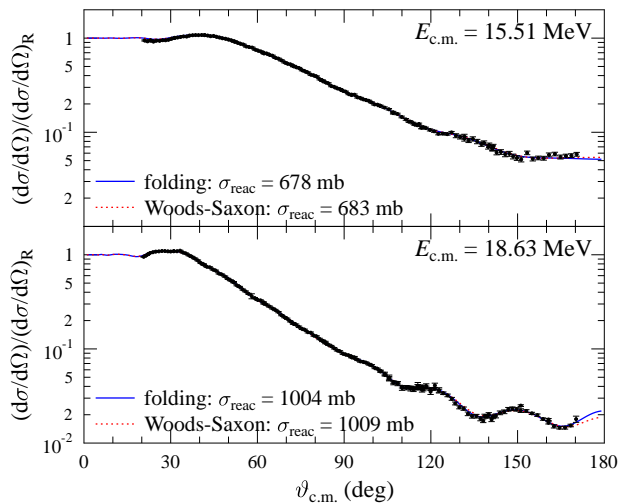}
\caption{
\label{fig:scat_tot}
(Color online)
Rutherford normalized elastic scattering cross sections of the
$^{89}$Y($\alpha,\alpha$)$^{89}$Y reaction at $E_{\rm{c.m.}}$ = 15.51\,MeV and
18.63\,MeV versus the angle in center-of-mass frame. The lines are calculated
from a folding potential (full blue line) and from a Woods-Saxon potential
(dotted red line) in the real part of the nuclear optical potential, and a
combined volume and surface Woods-Saxon potential in the imaginary part. Both
calculations accurately reproduce the angular distributions and thus provide
almost the same total reaction cross section \stot .
}
\end{figure}

Two experimental elastic scattering angular distributions of $^{89}$Y\raa
$^{89}$Y at energies of $E_{\rm{c.m.}} = 15.51$\,MeV and 18.63\,MeV
\cite{Kiss09} are shown in Fig.~\ref{fig:scat_tot} together with two different
fits to the data at each energy. The first fit considers a folding potential
in the real part and a combination of volume and surface Woods-Saxon
potentials in the imaginary part of the nuclear optical potential; it
is identical to the local potential in \cite{Kiss09}. In the second fit the
folding potential in the real part is replaced by a more flexible Woods-Saxon
potential of volume type where all Woods-Saxon parameters have been
readjusted. We find that both fits describe the data with similar $\chi^2$
values. The resulting total reaction cross sections \stot\ at 15.51\,MeV are
678\,mb in the case of considering a folding potential and 683\,mb for the 
Woods-Saxon potential and 1004\,mb (folding) and 1009\,mb
(Woods-Saxon) at 18.63\,MeV. The choice of the parameterization of the
real nuclear potential (folding or Woods-Saxon) has a minor influence on the
derived total reaction cross section of below 1\,\%. Changes in the shape of
the imaginary part also do not affect the extracted \stot\ significantly, as
long as the experimental data are precisely described. Predictions from global 
optical potentials typically agree with the \stot\ obtained from
experiments within 5\,\% in the $^{89}$Y case (for further details of 
global potentials for the $^{89}$Y-\al\ system, see discussion in
\cite{Kiss09}).

Because of the minor influence of the chosen potential parameterization we
list the total reaction cross sections \stot\ of $^{89}$Y \cite{Kiss09},
$^{92}$Mo \cite{Ful01}, $^{112,124}$Sn \cite{Gal05}, and $^{144}$Sm
\cite{Mohr97} from the original folding potential analyses of the given
papers. The results are summarized in Table \ref{tab:stot}.

\begin{table}[tbh]
\caption{\label{tab:stot}
Total reaction cross sections \stot\ derived from the \al -elastic angular
distributions of $^{89}$Y \cite{Kiss09}, $^{92}$Mo \cite{Ful01},
$^{112,124}$Sn \cite{Gal05}, and $^{144}$Sm \cite{Mohr97}.
}
\begin{center}
\begin{tabular}{ccr@{$\pm$}l}
\multicolumn{1}{c}{target} 
& \multicolumn{1}{c}{$E_{\rm{c.m.}}$ (MeV)}
& \multicolumn{2}{c}{\stot\ (mb) } \\
\hline
$^{89}$Y   & 15.51 &  678 & 20 \\
$^{89}$Y   & 18.63 & 1004 & 30 \\
$^{92}$Mo  & 13.20  &  238 & 7 \\
$^{92}$Mo  & 15.69 &  578 & 17 \\
$^{92}$Mo  & 18.62 &  883 & 26 \\
$^{112}$Sn & 13.90 &   93 & 3 \\
$^{112}$Sn & 18.84 &  695 & 21 \\
$^{124}$Sn & 18.90 &  760 & 23 \\
$^{144}$Sm & 19.45 &  404 & 12 \\
\end{tabular}
\end{center}
\end{table}

The absolute normalization of the experimental angular distributions is an
important prerequisite for the accurate determination of the total cross
section \stot . Fortunately, it can be taken from data at very forward
angles where the scattering cross section is given by the Rutherford cross
section. The smallest measured angle has
to reach the range where the angular distribution does not deviate from the
Rutherford cross section by more than 1\,\%. The huge Rutherford cross
sections at forward angles typically lead to high count rates, and thus a
careful deatime correction is required. Furthermore, the steep angular
dependence of the Rutherford cross section requires a precise angular
calibration. These requirements are fulfilled by our experiments, and the
resulting uncertainty in the absolute normalization of our experimental 
data is of the order of 1\,\%; for details,
see Refs.~\cite{Mohr97,Ful01,Gal05,Kiss09}.

The data at backward angles strongly affect 
the imaginary part of the potential which in turn determines the deviations of
the reflexion coefficients $\eta_L$ from unity. (Obviously, $\eta_L = 1$ for
all angular momenta $L$ and thus \stot = 0 is found for all nuclear potentials
without imaginary part.) Typical uncertainties of individual data points at
backward angles 
remain below $3-4$\,\% (mainly defined by statistics). Together with the huge
number of data points in \cite{Mohr97,Ful01,Gal05,Kiss09} the shape of the
angular distribution is well-defined within 2\,\% over the full measured
angular range. In a simple view, the total reaction cross section \stot\ is
approximately given by the deviation of the measured angular distribution from
the Rutherford cross section, and the uncertainty of the backward angular
distribution enters directly into the uncertainty of \stot .

Combining the above uncertainties of less than 1\,\% from the choice of the
potential, about 1\,\% from the absolute normalization at forward angles,
about 2\,\% from the backward cross sections, and 1\,\% from the not measured
angular range above $170^\circ$, this leads to a total uncertainty of about
3\,\% for \stot\ from the analysis of the scattering data of
\cite{Mohr97,Ful01,Gal05,Kiss09} using error quadratic summing.

\begin{figure}
\includegraphics[bbllx=60,bblly=40,bburx=355,bbury=290,width=\columnwidth,clip=]{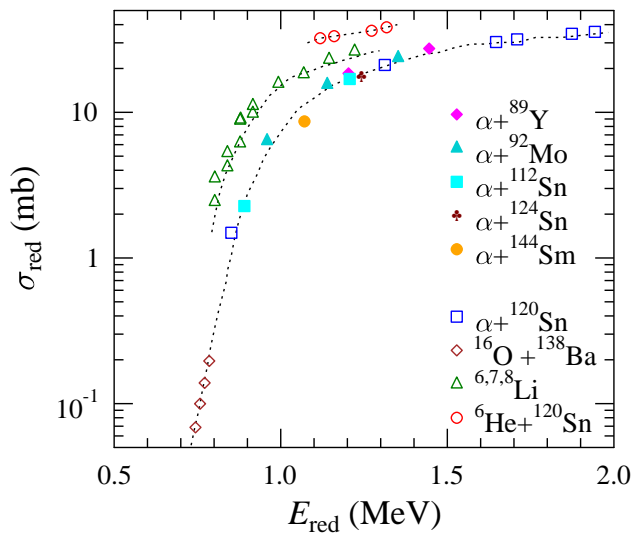}
\caption{
\label{fig:sig_red}
(Color online)
Reduced reaction cross sections $\sigma_{\rm{red}}$ versus reduced energy
$E_{\rm{red}}$ for tightly bound \al -particles and $^{16}$O, weakly bound
$^{6,7,8}$Li projectiles, and exotic $^{6}$He. (Update of Fig.~4 from
\cite{Far10} with additional data from \cite{Mohr10}). The error bars of the
new data (full data points) are omitted because they are smaller than the
point size. The lines are to guide the eye.
}
\end{figure}

Finally, in Fig.~\ref{fig:sig_red} we compare the new total reaction cross
sections  of \al -induced 
reactions to a series of data for exotic nuclei ($^6$He), weakly bound nuclei
($^{6,7,8}$Li), and tightly bound particles (\al\ or $^{16}$O); this is
basically an update of Fig.~4 of \cite{Far10} with new additional data
points from this study and from $^{120}$Sn \cite{Mohr10}. For this comparison
we use the so-called reduced energy $E_{\rm{red}}$ and reduced reaction cross
section $\sigma_{\rm{red}}$ which are given by
\begin{eqnarray}
E_{\rm{red}} & = & \frac{\bigl(A_P^{1/3}+A_T^{1/3}\bigr) E_{\rm{c.m.}}}{Z_P Z_T} \\
\sigma_{\rm{red}} & = & \frac{\sigma_{\rm{reac}}}{\bigl(A_P^{1/3}+A_T^{1/3}\bigr)^2}
\label{eq:red}
\end{eqnarray}
The reduced
energy $E_{\rm{red}}$ takes into account the different heights of the Coulomb
barrier in the systems under consideration, whereas the reduced reaction cross
section \sred\ scales the measured total reaction cross section \stot\
according to the geometrical size of the projectile-plus-target system.
A smooth behavior for all \sred\ of \al -induced reactions is found. The
obtained 
values for \sred\ are smaller for tightly bound projectiles (\al , $^{16}$O)
compared to weakly bound projectiles ($^{6,7,8}$Li) and halo projectiles
($^6$He). The lines in Fig.~\ref{fig:sig_red} are shown to guide the eye
(taken from \cite{Far10}). A deeper theoretical understanding of these
different energy dependencies will be helpful for the prediction of reaction
cross sections within the statistical model. 

In conclusion, we have extracted total reaction cross sections \stot\ from our
previously published experimental \al\ scattering data on $^{89}$Y,
$^{92}$Mo, $^{112,124}$Sn, and $^{144}$Sm
\cite{Mohr97,Ful01,Gal05,Kiss09}. The derived cross sections \stot\ have small
uncertainties of about 3\,\% and show a smooth and systematic energy
dependence which is consistent with other \al -scattering results and data for
another tightly bound projectile $^{16}$O. The energy dependence of the
derived reduced cross section \sred\ significantly differs from reactions
induced by weakly bound and exotic projectiles with halo properties like
e.g.\ $^6$He.

\begin{acknowledgments}
We thank R.\ Lichtenth\"aler for encouraging discussions and providing the
data of their Fig.~4 of Ref.~\cite{Far10}. This work is part of the
{\sc{EuroGENESIS}} project and was supported by OTKA (K68801, NN83261) and 
European Research Council grant agreement no.~203175.
\end{acknowledgments}

\end{document}